\documentclass[aps,prb,reprint,showpacs]{revtex4-1}
\usepackage{amsmath}
\usepackage{natbib}
\usepackage{graphicx}
\usepackage{hyperref}
\usepackage[all]{hypcap}
\usepackage{microtype}

\begin{document}
\title{Pattern formation in the dipolar Ising model on a two-dimensional honeycomb lattice}

\author{Robert R\"uger}
\author{Roser Valent\'\i}
\affiliation{Institut f\"ur Theoretische Physik, Goethe-Universit\"at Frankfurt, Max-von-Laue-Stra{\ss}e 1, 60438 Frankfurt am Main, Germany}

\date{\today}

\begin{abstract}
We present Monte Carlo simulation results for a two-dimensional Ising model with ferromagnetic nearest-neighbor couplings and a competing long-range dipolar interaction on a honeycomb lattice. Both structural and thermodynamic properties are very similar to the case of a square lattice, with the exception that structures reflect the sixfold rotational symmetry of the underlying honeycomb lattice. To deal with the long-range nature of the dipolar interaction we also present a simple method of evaluating effective interaction coefficients, which can be regarded as a more straightforward alternative to the prevalent Ewald summation techniques.
\end{abstract}


\maketitle

\section{Introduction}

The two-dimensional Ising model with ferromagnetic nearest-neighbor interactions and long-range antiferromagnetic interactions is probably the simplest model system for the formation of magnetic domains, for instance see T. Garel and S. Doniach.~\cite{PhysRevB.26.325} Physical systems for which the application of such a model is justified are, for example, ultrathin metal films on metal substrates, provided that there is a strong magnetocrystalline anisotropy which favours spin alignment perpendicular to the plane of the film.~\cite{RevModPhys.72.225} In addition to the nearest-neighbor coupling via exchange interaction, the magnetic dipole-dipole interaction then realizes a long-range antiferromagnetic coupling that decreases as~$r^{-3}$. A thin magnetic film may therefore be described by the following Hamiltonian:
\begin{equation}\label{e:hamiltonian_1}
H = - 2 J \sum_{\langle i,j \rangle} S^z_i S^z_j - g \sum_{\substack{i,j \\ i \neq j}} \frac{S^z_i S^z_j}{\left|\vec r_i - \vec r_j\right|^3 } - B^z \sum_i S^z_i ,
\end{equation}
where $S^z_i$ is the $z$~component of a spin~$1/2$ operator on site~$i$, $J > 0$ corresponds to the ferromagnetic exchange interaction constant, $g$ defines the strength of the dipolar coupling and $B^z$ denotes an external magnetic field oriented along the $z$~direction. Without loss of generality, we measure distances~$r$ in units of the nearest-neighbor distance. Since the dipolar interaction is inherently antiferromagnetic~($g<0$) and a purely dipolar Ising model~($J=0$) has the usual antiferromagnetic ground state, an additional antiferromagnetic exchange interaction would simply increase the transition temperature.~\cite{PhysRevLett.75.950} Therefore, only the case of a ferromagnetic exchange interaction~($J>0$) is interesting. We also want to assume for the remainder of this article that the exchange interaction is strong enough so that it is the dominant coupling between nearest neighbors, therefore~$J > |g|$.

As metal-on-metal films have many technological applications, including for example electronics, data storage, and catalysis,~\cite{RevModPhys.72.225} extensive studies have been performed to investigate the properties of Hamiltonian~\eqref{e:hamiltonian_1}. Most of the metal-on-metal films happen to be square or triangular lattice systems, so these studies naturally have had their focus on square~\cite{PhysRevB.51.16033, PhysRevLett.75.950, PhysRevB.46.6387, PhysRevB.73.184425, PhysRevB.76.054438} and triangular~\cite{ActaPhysPolA.113.2.951} lattices. Two-dimensional magnets with other lattices are less obvious, but recently the honeycomb lattice has been discussed~\cite{PhysRevB.84.201104, NatCommun.2.596} in the context of the $(111)$~bilayer of LaNiO$_3$, which shows a very rich magnetic phase diagram. To the best of our knowledge there are no publications dealing with Hamiltonian~\eqref{e:hamiltonian_1} on an underlying honeycomb lattice. This article presents Monte Carlo simulation results for the thermodynamic and structural properties of such a system.

Let us briefly recall the existing results for the square lattice (for more extensive reviews, see Refs.~\onlinecite{RevModPhys.72.225} and~\onlinecite{RecResDev.5.II.751}): It has been analytically established~\cite{PhysRevB.51.16033} that the ground state shows a striped pattern of width~$h$, where~$h$ increases exponentially with the relative strength~$J/|g|$ of the exchange interaction in the limit~$J \gg |g|$. This implies that even an infinitesimal antiferromagnetic dipolar interaction will destroy the spontaneous magnetization of the purely ferromagnetic ground state. Intermediate between the low-temperature striped phase and the high-temperature paramagnetic phase with maximum entropy, a third phase is found which shows well defined magnetic domains that form mazelike patterns.~\cite{PhysRevLett.75.950} This phase has been  called the tetragonal phase due to the predominantly rectangular corners of those domains. Numerical calculations of the structure factor
\begin{equation}\label{e:structure_factor}
\sigma(\vec k) = \left\langle \left| \sum_j S^z_j \exp\left( \mathrm{i} \, \vec k \cdot \vec r_j \right) \right|^2 \right\rangle ,
\end{equation}
where $\vec{k}$ is a wavevector in  reciprocal space, show the fourfold rotational symmetry of the tetragonal phase.~\cite{PhysRevB.51.16033} The transition from the striped phase to the tetragonal phase is accompanied by a sharp peak in the specific heat at temperatures below the Shottky anomaly shoulder, which itself can be associated with the tetragonal-paramagnetic transition.~\cite{PhysRevLett.75.950} Larger values of~$J/|g|$ generally increase all temperatures and make the striped-tetragonal peak less pronounced relative to the Shottky shoulder.~\cite{PhysRevLett.75.950} More recent studies~\cite{PhysRevB.73.184425, PhysRevB.76.054438} have revealed the existence of an intermediate nematic phase between the striped phase and the tetragonal phase that is only stable for very specific values of~$J/|g|$ close to some of the ground state stripe width transitions.

In the present work we examine the domain pattern formation on a  honeycomb lattice in the framework of the dipolar Ising model and compare the obtained results to the square lattice case. In particular we want to determine whether a change of the underlying lattice only modifies transition temperatures or also changes the system's behavior qualitatively. For this purpose, we present an efficient method to numerically evaluate effective interaction coefficients. This method can be regarded as a simple alternative to the known Ewald summation techniques.

\section{Methods}

Monte Carlo simulations of spin systems with long-range interactions are inherently difficult. The reason for that lies in the nontrivial implementation of periodic boundary conditions. Also, the  $\mathcal O (N^2)$ number of couplings, where $N$ denotes the number of spins, make the evaluation of energy differences computationally expensive. In the following we will describe the methods we used to obtain our results.

\begin{figure}[b!]
\includegraphics[width=0.5\columnwidth]{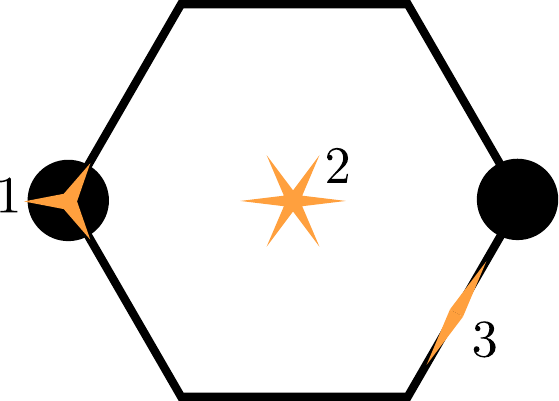}
\caption{\label{f:rotsym}(Color online) Rotational symmetries of the honeycomb lattice: threefold symmetry for rotations around one of the basis atoms (point~$1$), sixfold rotational symmetry around the center of the unit cell~(point $2$) and twofold symmetry around the center of nearest-neighbor connections (point 3).}
\end{figure}

As the tetragonal and striped phases that occur on a square lattice clearly reflect the rotational symmetries of the underlying lattice, special care should be taken in case of the honeycomb lattice to ensure that the shape and boundary conditions of the finite system allow the formation of structures with the expected rotational symmetries. Fig.~\ref{f:rotsym} illustrates the three rotational symmetries of the honeycomb lattice.

Finite size effects can be reduced through the use of periodic boundary conditions. For systems with long-range interactions, periodic boundary conditions can be implemented by tiling the entire space with replicas of the original finite system.~\cite{PhysRevB.46.6387} Note that this is equivalent to the treatment of an infinite system, where only states with certain translational invariances are considered.

\begin{figure}[tbp]
\includegraphics[width=\columnwidth]{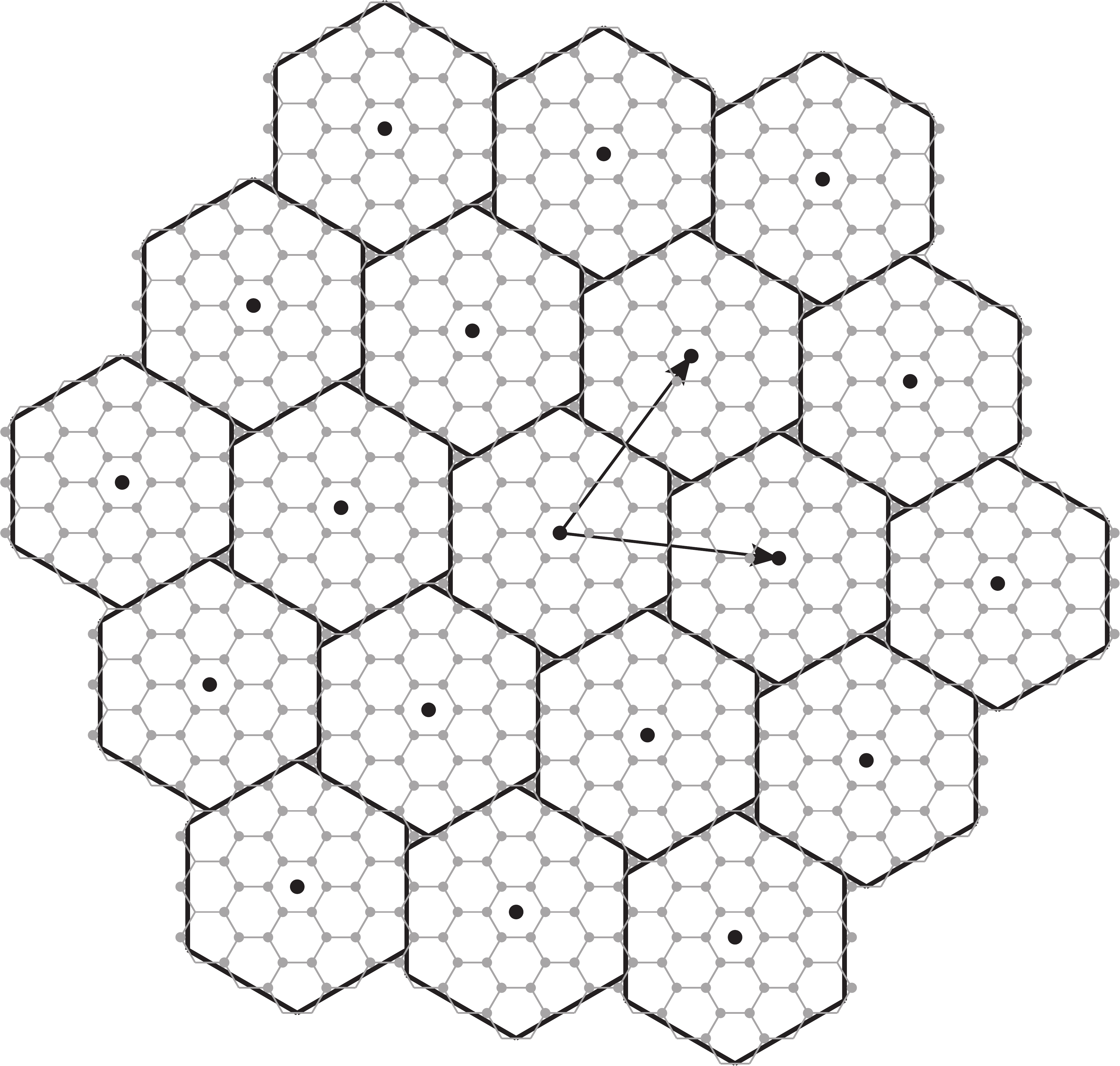}
\caption{\label{f:boundary}Periodic boundary conditions on a honeycomb lattice. The original finite system's shape is approximated by the central black hexagon. Its replicas form a triangular Bravais lattice, whose basis vectors are marked in black. According to our nomenclature this would be a $(3,3)$~aggregation.}
\end{figure}

Both the rotational symmetries and the requirement of the system to be properly tileable are satisfied if one chooses the simulated system's shape to be a regular hexagon. This hexagon unit as well as the hexagons' tiling  is illustrated in Fig.~\ref{f:boundary}.
Let us introduce the name ``aggregation'' for the combination of the original system and all of its replicas. Note the equivalence between the tiling of the system's replicas to form the aggregation and the tiling of the unit cells within the original system. It is convenient to use the side length of the original system in number of unit cells as a measure for the size~$n$ of the simulated system. Since the aggregation itself is also a regular hexagon, it is then straightforward to measure its size~$m$ as the length of its sides in units of replicas. In this way, the entire size and shape of the aggregation is specified by a pair~$(n,m)$ of integers. Note that the original system as well as the entire aggregation are sixfold rotationally symmetric around their center, which implies threefold and twofold symmetry.

Now that we have reduced the infinitely large system to a finite system with replicas, we can make use of this new periodicity. Ignoring for the moment the coupling to an external magnetic field, we rewrite the Hamiltonian~\eqref{e:hamiltonian_1} by using $S^z_i \equiv S^z(\vec r_i) = S^z(\vec R + \vec r_i)$ where $\vec R$ is the translation from the original system to any of the replicas:
\begin{subequations}
\begin{equation}
H = - \sum_{i,j} \sum_{\vec R} \left[ \Lambda( \vec r_i, \vec R + \vec r_j ) + \Gamma( \vec r_i, \vec R + \vec r_j ) \right] S^z_i S^z_j .
\end{equation}
Here the function~$\Lambda$ is a nonzero constant for nearest neighbors~(NN) only, whereas $\Gamma$ takes care of the dipole-dipole interaction without the unphysical interaction of a spin with itself. Note, however, that a spin does interact with its own copies in the replicated systems at~$\vec R$:
\begin{align}
\Lambda( \vec r_i, \vec R + \vec r_j ) &= \begin{cases} J & \text{if $\vec r_i$ and $\vec R + \vec r_j$ are NN,} \\ 0 & \text{otherwise,} \end{cases} \displaybreak[0]\\ 
\Gamma( \vec r_i, \vec R + \vec r_j ) &= \begin{cases} 0 & \text{if $\vec r_i = \vec R + \vec r_j$,} \\ \frac{g}{\left|\vec r_i - \vec R - \vec r_j\right|^3} & \text{otherwise.} \end{cases}
\end{align}
\end{subequations}
As the sum over all replicas depends only on the indices~$i$ and~$j$, but not on the orientation of those spins, it can be calculated in advance, which reduces the Hamiltonian to the general Ising model Hamiltonian with effective interaction coefficients:
\begin{subequations}
\begin{align}
H &= - \sum_{i,j} J_{ij}^\text{eff} S^z_i S^z_j - B^z \sum_i S^z_i , \displaybreak[0]\\
J_{ij}^\text{eff} &= \sum_{\vec R} \left[ \Lambda( \vec r_i, \vec R + \vec r_j ) + \Gamma( \vec r_i, \vec R + \vec r_j ) \right] .
\end{align}
\end{subequations}
In addition to the exchange interaction, if~$S^z_i$ and~$S^z_j$ (or any of its copies) are nearest neighbors, this effective interaction coefficient~$J_{ij}^\text{eff}$ also includes the dipole-dipole interaction of $S^z_i$ with $S^z_j$ and all copies of $S^z_j$.

It is obvious that, once the effective interaction coefficients have been calculated, the time required to perform a Monte Carlo step for a $(n,m)$~aggregation will only depend on the number of spins and therefore on the size~$n$ of the original system. The accuracy of the effective interaction coefficients depends on the size~$m$ of the aggregation though, but since the~$J_{ij}^\text{eff}$ only have to be calculated once, the aggregation size~$m$ can be quite large. Note that an increase in the system size~$n$ for constant values of~$m$ also leads to an increase in the accuracy of the effective interaction coefficients, as it corresponds to larger absolute values of the translation vectors~$\vec R$ and therefore increases the distance at which the sum is truncated.

Due to the finite size of the aggregation, the dipolar part of the effective interaction strength will systematically be underestimated. Assuming a $(n,m)$~aggregation, one can try to compensate for this by calculating the effective interaction~$J_{ii}^\text{eff}$ of a spin with its own copies for a $(n,m')$~aggregation, where $m' \gg m$. As $S^z_i$ will certainly not be its own nearest neighbor, the coefficient~$J_{ii}^\text{eff}$ contains only dipolar interactions. In the limit $m' \rightarrow \infty$, the difference between~$J_{ii}^\text{eff}$ for~$(n,m)$ and $(n,m')$~aggregations will just be the remaining dipole interaction that has been neglected with the $(n,m)$~aggregation. One can now add this difference to all effective interaction coefficients to compensate approximately for the systematic underestimation. Note that this correction, which has been calculated for the interaction~$J_{ii}^\text{eff}$ with the spin's own copies, is added to all other coefficients, including~$J_{ij}^\text{eff}$ with~$i \neq j$. It therefore also accounts approximately for the interaction with copies of all other spins. This approximation is justified by the fact that the dipolar potential is almost flat for very large~$r$ and therefore not sensitive to the exact relative positions. Because the actual calculation with the larger $(n,m')$~aggregation only has to be performed for a single spin and its own copies, the computational effort is negligible.

A significant speedup in the calculation of the effective interaction coefficients can be achieved if one manages to exploit symmetries to reduce the number of coefficients that have to be calculated. As only relative positions matter, it is quite obvious that not all of the~$N^2$ coefficients in a system with $N$~spins will actually be different. In principle, one could also reduce the memory footprint of the coefficient table in this way, but one would have to evaluate very carefully if the overhead of selecting the right coefficient does not slow down the Monte Carlo step significantly.

This direct calculation of the effective interaction coefficients is, compared to the alternative Ewald summation techniques,~\cite{ANDP:ANDP19213690304} a very simple and straightforward method. For this particular system it works very well, since it is a two-dimensional system where the long-range interaction decreases with~$r^{-3}$. This dependence makes the series unproblematically convergent. Using the proposed methods, we were able to replicate the known results for the square lattice within the statistical uncertainty. All of our honeycomb lattice simulations were performed on a $(24,24)$ aggregation, which corresponds to a system of $N=3314$ spins. A $(24,1000)$ aggregation was used to compensate for the underestimation.  This calculation resulted in a relative increase in the effective coefficients of about~$10^{-2}$ for couplings of spins that are far away from each other, and about~$10^{-6}$ for nearest neighbors.

Metropolis dynamics~\cite{1953JChPh..21.1087M} with uniformly distributed single spin-flip attempts were used to evolve the system. Markov chain correlation was dealt with using binning and bootstrapping~\cite{Efr79} techniques. The energy and temperature scale is defined by measuring~$J$ and~$k_B T$ in units of~$-g$.

\section{Results}

In order to perform a Monte Carlo simulation over the whole temperature range, one first has to determine which states have the lowest energies. Initializing the equilibration period of a low temperature simulation with a state that is typical for high temperatures  will not correctly equilibrate the system: An almost steepest descent in energy will most likely trap the system in a local minimum from which no physical information can be extracted. Therefore we have first performed a simulated annealing to relax the system to a low energy state, which was then used to initialize the equilibration period of the simulations.

\begin{figure}[tbp]
\includegraphics[width=0.45\columnwidth]{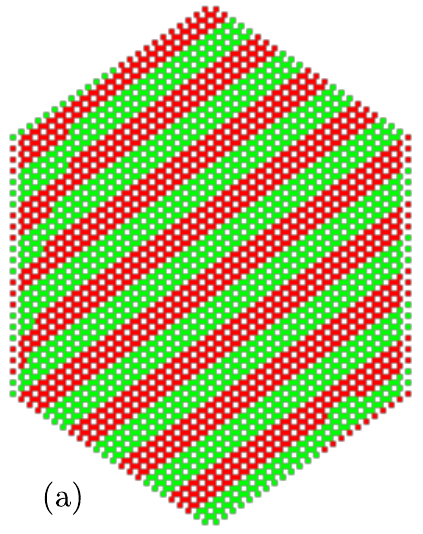}
\hskip 0.05\columnwidth
\includegraphics[width=0.45\columnwidth]{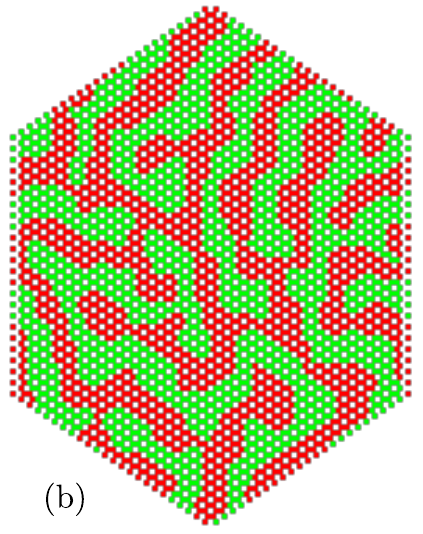}
\caption{\label{f:structures}(Color online) (a) Striped  phase for~$J = 6$ at~$k_B T = 0.4$ and (b) hexagonal phase
for $J = 6$ at $k_B T = 1.25$.}
\end{figure}

We have found the simulated annealing to result in a striped state if the system's temperature is decreased linearly from~$T=5$ to~$0$ over the course of $10^7$~Monte Carlo steps. This indicates that the ground state of the dipolar Ising model on the honeycomb lattice is indeed striped, similar to the state that is plotted in Fig.~\ref{f:structures}(a).
Note that one has to be very careful to choose the system size to be compatible with the width of the stripes that the system would like to form, in order not to introduce an artificial frustration. As the width of the stripes is determined by the relative strength of exchange and dipolar interaction, one can also adjust~$J$ to make the stripe width compatible with the given system size. We have found $J=6$ to result in a stripe width that is compatible with the system size of~$n=24$ which was used in all our simulations. Naturally, the stripes become wider if the strength of the ferromagnetic exchange interaction is increased.

Having determined the low-temperature states to be striped by simulated annealing, we have equilibrated the system at constant temperatures for $2 \times 10^5$~Monte Carlo steps and recorded thermodynamic and structural properties for $8 \times 10^6$ time steps. We find that the low-temperature striped phase is followed by a phase with a complex domain structure which is plotted in Fig.~\ref{f:structures}(b). 
\begin{figure}[tbp]
\includegraphics[width=\columnwidth]{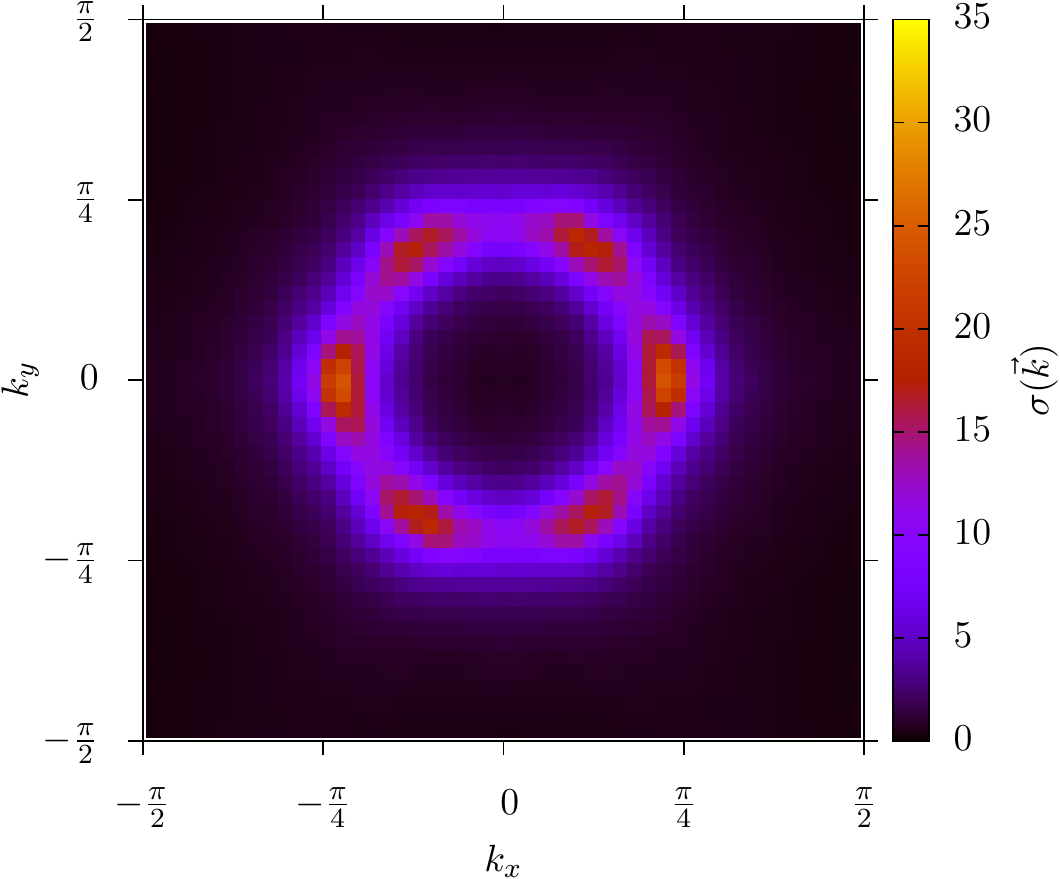}
\caption{\label{f:J6T125_sfplot}(Color online) Structure factor~$\sigma(\vec k)$  in the hexagonal phase at $k_B T = 1.25$ and~$J = 6$.}
\end{figure}
We call this phase hexagonal, due to the sixfold rotational symmetry that is visible in the structure factor~$\sigma(\vec k)$ plotted in Fig.~\ref{f:J6T125_sfplot}. The six-peaked shape of the structure factor remains unchanged if the sum in Eq.~\eqref{e:structure_factor} is restricted to one of the two basis atoms in the honeycomb unit cell. We can conclude from this that the hexagonal phase shows the rotational symmetry of the triangular Bravais lattice that underlies the honeycomb lattice. It is interesting to note that a change of the lattice from square to honeycomb changes only the intermediate-temperature phase from tetragonal to hexagonal, while the low-temperature phase is striped for both lattices.

We do not observe the three-peaked structure of the energy histrogram that was used by Cannas et al.\cite{PhysRevB.73.184425} to identify the square lattice's nematic phase. We therefore have to conclude that there is no nematic phase for the particular value of~$J/|g|$ used in our calculations.

\begin{figure}[tbp]
\includegraphics[width=\columnwidth]{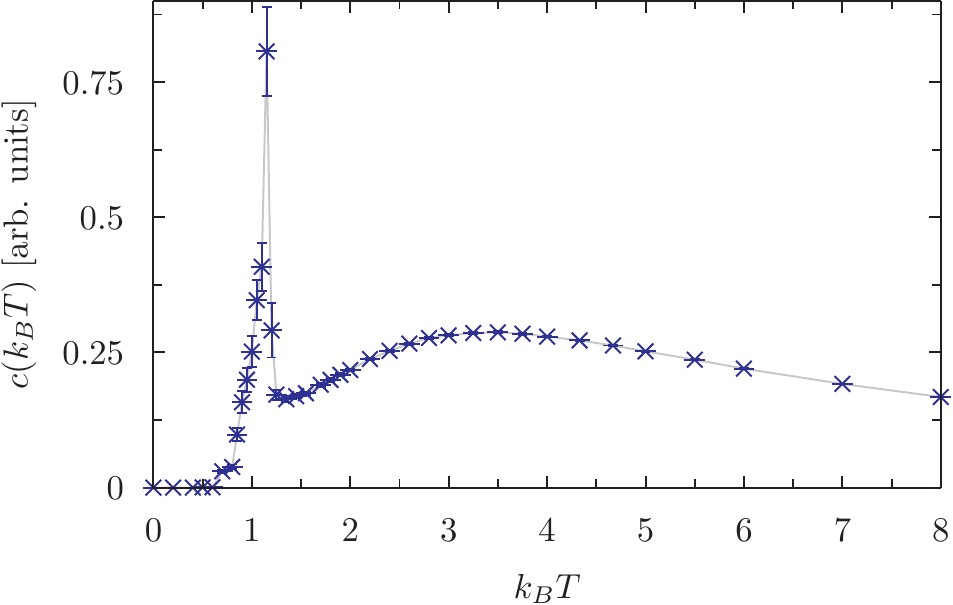}
\caption{\label{f:J6_cplot}(Color online) Specific heat capacity per spin~$c(k_B T)$ as a function of temperature for a relative interaction strength of~$J = 6$. The line is just a guide to the eye.}
\end{figure}

Similar to the striped-tetragonal transition on the square lattice, the striped-hexagonal transition manifests itself in a sharp peak in the specific heat capacity at temperatures below the expected Shottky anomaly shoulder as shown in Fig.~\ref{f:J6_cplot}.
We expect the striped-hexagonal peak to become less pronounced for larger~$J$, as is the case on a square lattice.

In summary, by means of Monte Carlo simulations we have investigated the temperature dependence of domain pattern formation in a dipolar Ising model on a two-dimensional honeycomb lattice. Taking into account the symmetries of the honeycomb lattice, we have reduced the infinite system to a finite system with replicas, which can be described by the general Ising model Hamiltonian with effective interaction coefficients. We have presented a straightforward method of calculating the effective interaction coefficients and a procedure to compensate approximately for their systematic underestimation. We find that the honeycomb lattice shows two distinct phase transitions: from a striped phase at low temperatures via a hexagonal phase at intermediate temperatures to a disordered phase at high temperatures. Both transitions are associated with maxima in the specific heat. While the thermodynamic properties of the honeycomb lattice system are found to be virtually identical to its square lattice counterpart, the emerging patterns clearly reflect the different rotational symmetry of the underlying lattice.

Future work should focus on obtaining the entire phase diagram of the model in order to determine the values of~$J/|g|$ at which the ground state shows transitions in stripe width. For those values one can then systematically search for the honeycomb lattice's analogon to the nematic phase~\cite{PhysRevB.73.184425, PhysRevB.76.054438} in between the striped and the hexagonal phases.

{\it Acknowledgements.--} We would like to thank Oleg Tcher\-ny\-shy\-ov for providing useful comments and the Deutsche Forschungsgemeinschaft for financial support through Grant No. SFB/TR 49.

%

\end{document}